\documentclass[sigconf]{acmart}
\AtBeginDocument{%
  }

\setcopyright{acmlicensed}
\copyrightyear{2018}
\acmYear{2018}
\acmDOI{XXXXXXX.XXXXXXX}
\acmConference[Conference acronym 'XX]{Make sure to enter the correct
  conference title from your rights confirmation email}{June 03--05,
  2018}{Woodstock, NY}
\acmISBN{978-1-4503-XXXX-X/2018/06}

\usepackage{xspace}
\usepackage{xcolor}
\usepackage{subcaption}
\usepackage{stfloats}

\newcommand{\model}{WHALE\xspace}
\newcommand{\layer}{WHALE Layer\xspace}

\definecolor{renqinbrown}{RGB}{150,75,0}

\copyrightyear{2026}
\acmYear{2026}
\setcopyright{cc}
\setcctype{by-nc-nd}
\acmConference[RecSys '26]{20th ACM Conference on Recommender Systems}{September 27-October 02, 2026}{Minneapolis, MN, USA}
\acmBooktitle{20th ACM Conference on Recommender Systems (RecSys '26), September 27-October 02, 2026, Minneapolis, MN, USA}
\acmDOI{10.1145/3773078.3831927}
\acmISBN{979-8-4007-2284-4/2026/09}

\begin{document}

\title{WHALE: A Scalable Unified Model for Recommendation with Wukong-HSTU Architecture}

\author{Renqin Cai}
\authornote{Both authors contributed equally to this work.}
\affiliation{%
  \institution{Meta Platforms, Inc.}
  \country{USA}}
\email{renqincai@meta.com}

\author{Dawei Sun}
\authornotemark[1]
\affiliation{%
  \institution{Meta Platforms, Inc.}
  \country{USA}}
\email{dwsun@meta.com}

\author{Yuanjun Yao}
\affiliation{%
  \institution{Meta Platforms, Inc.}
  \country{USA}}
\email{yjyao@meta.com}

\author{Zhiyong Wang}
\affiliation{%
  \institution{Meta Platforms, Inc.}
  \country{USA}}
\email{zhiywang@meta.com}

\author{Velvin Fu}
\affiliation{%
  \institution{Meta Platforms, Inc.}
  \country{USA}}
\email{velvinfu@meta.com}

\author{Maggie Zhuang}
\affiliation{%
  \institution{Meta Platforms, Inc.}
  \country{USA}}
\email{magicmag@meta.com}

\author{Yu Shi}
\affiliation{%
  \institution{Meta Platforms, Inc.}
  \country{USA}}
\email{yushi2@meta.com}

\author{Zhongnan Fang}
\affiliation{%
  \institution{Meta Platforms, Inc.}
  \country{USA}}
\email{zhongnan@meta.com}

\author{Xuan Cao}
\affiliation{%
  \institution{Meta Platforms, Inc.}
  \country{USA}}
\email{xuancao@meta.com}

\author{Jing Qian}
\affiliation{%
  \institution{Meta Platforms, Inc.}
  \country{USA}}
\email{jingqian@meta.com}

\author{Rui Li}
\affiliation{%
  \institution{Meta Platforms, Inc.}
  \country{USA}}
\email{ruili@meta.com}

\renewcommand{\shortauthors}{Cai et al.}

\begin{abstract}
  As scalability becomes increasingly important in recommendation model,
  recent architectures have advanced the modeling of two broad sources of
  ranking signals along separate paths: non-sequence features, including user,
  item, context, and cross features; and sequence features from user behavior
  histories. Wukong and HSTU
  have emerged as representative scalable backbones for these paths:
  Wukong scales high-order non-sequence
  feature-interaction modeling, while HSTU scales long user-behavior
  sequence modeling. Despite their complementary strengths, practical
  architectures that combine these two types of feature modeling remain
  underexplored. We present \model, a scalable unified recommendation
  architecture that jointly models non-sequence and sequence features on top
  of Wukong and HSTU. Each \model layer contains a Wukong module, an HSTU
  module, and an attention-based fusion module in which Wukong-derived
  interaction representations query HSTU-derived behavior representations. This
  design keeps both backbones active throughout the network and enables
  progressive Wukong--HSTU exchange, allowing high-order feature crosses
  to repeatedly retrieve fine-grained evidence from long user histories.
  To make \model practical for industrial deployment,
  we introduce customized Triton kernels and other model-systems co-design
  techniques to improve training and inference efficiency. On large-scale industrial recommendation data, \model achieves consistent gains in offline experiments. Additionally, it delivers positive
online gains with a modest serving-throughput trade-off. The method has been deployed in production systems. Overall,
WHALE provides a practical example of how these two sources of
information can be scalably unified in an industrial recommendation model.
\end{abstract}



%%
%% Keywords. The author(s) should pick words that accurately describe
%% the work being presented. Separate the keywords with commas.

\maketitle

\begin{figure*}[t]
\centering
\begin{subfigure}[t]{0.27\textwidth}
\centering
\includegraphics[width=\linewidth]{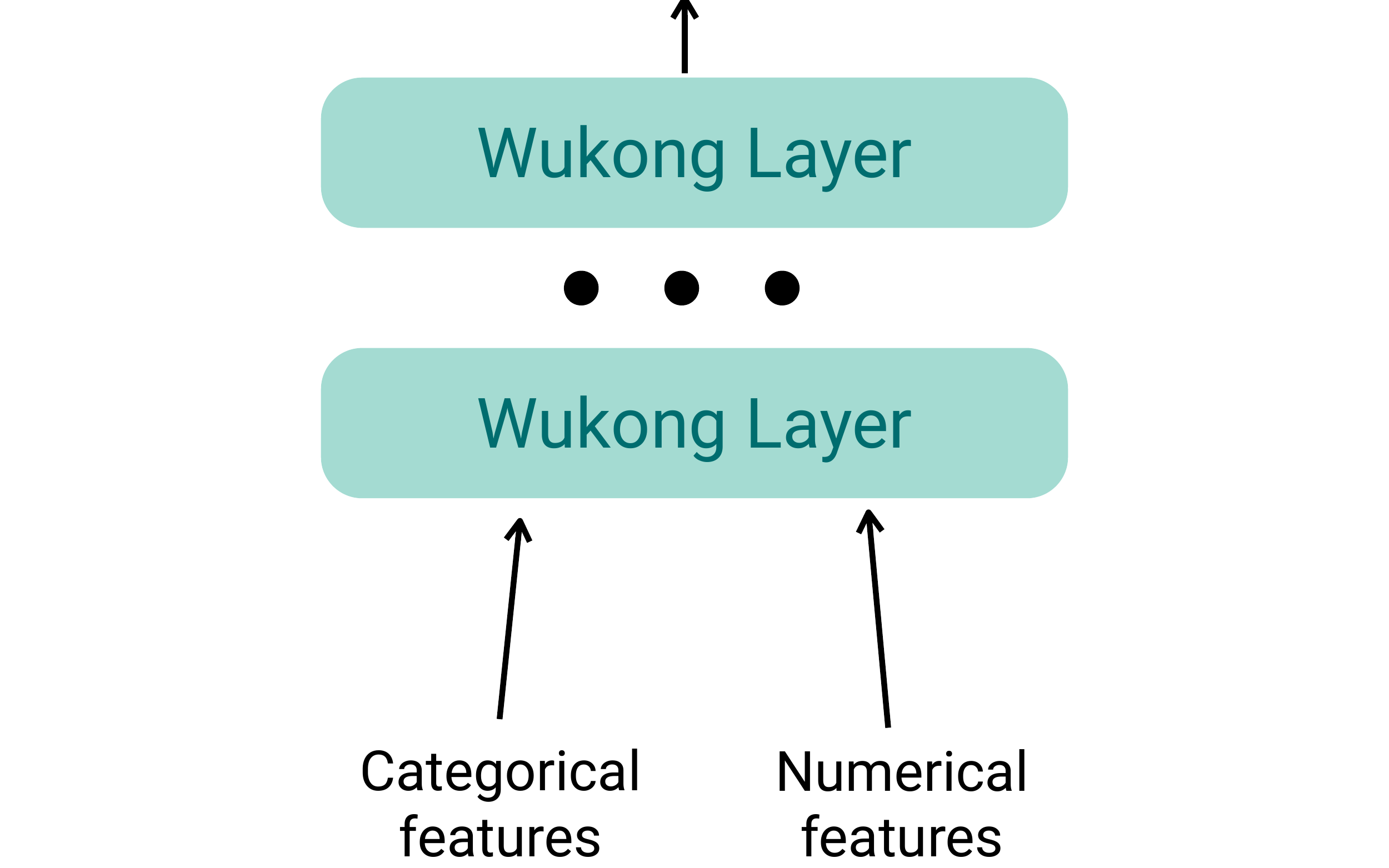}
\caption{Wukong architecture.}
\label{fig:wukong-architecture}
\end{subfigure}\hfill
\begin{subfigure}[t]{0.18\textwidth}
\centering
\includegraphics[width=\linewidth]{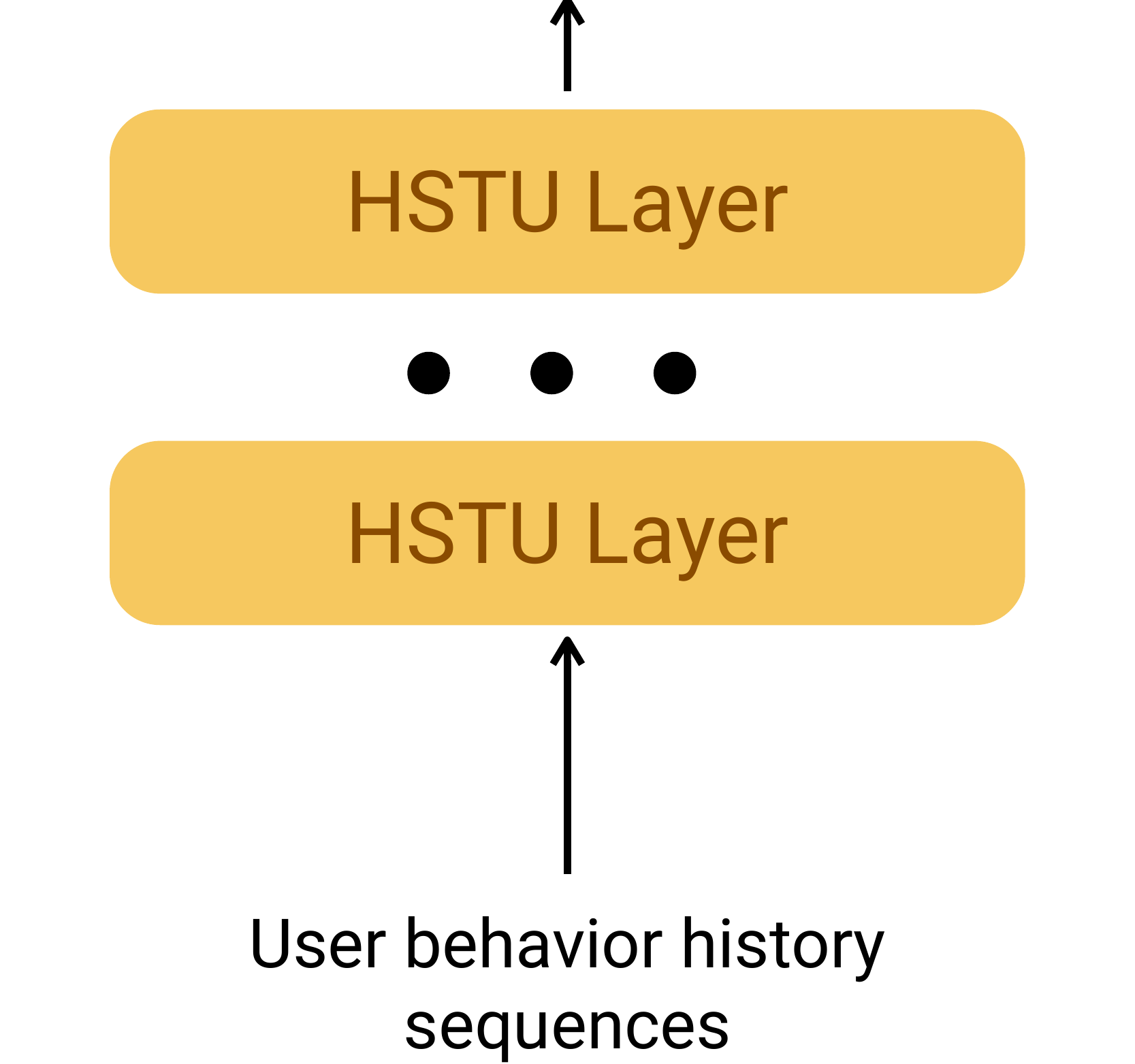}
\caption{HSTU architecture.}
\label{fig:hstu-architecture}
\end{subfigure}\hfill
\begin{subfigure}[t]{0.27\textwidth}
\centering
\includegraphics[width=\linewidth]{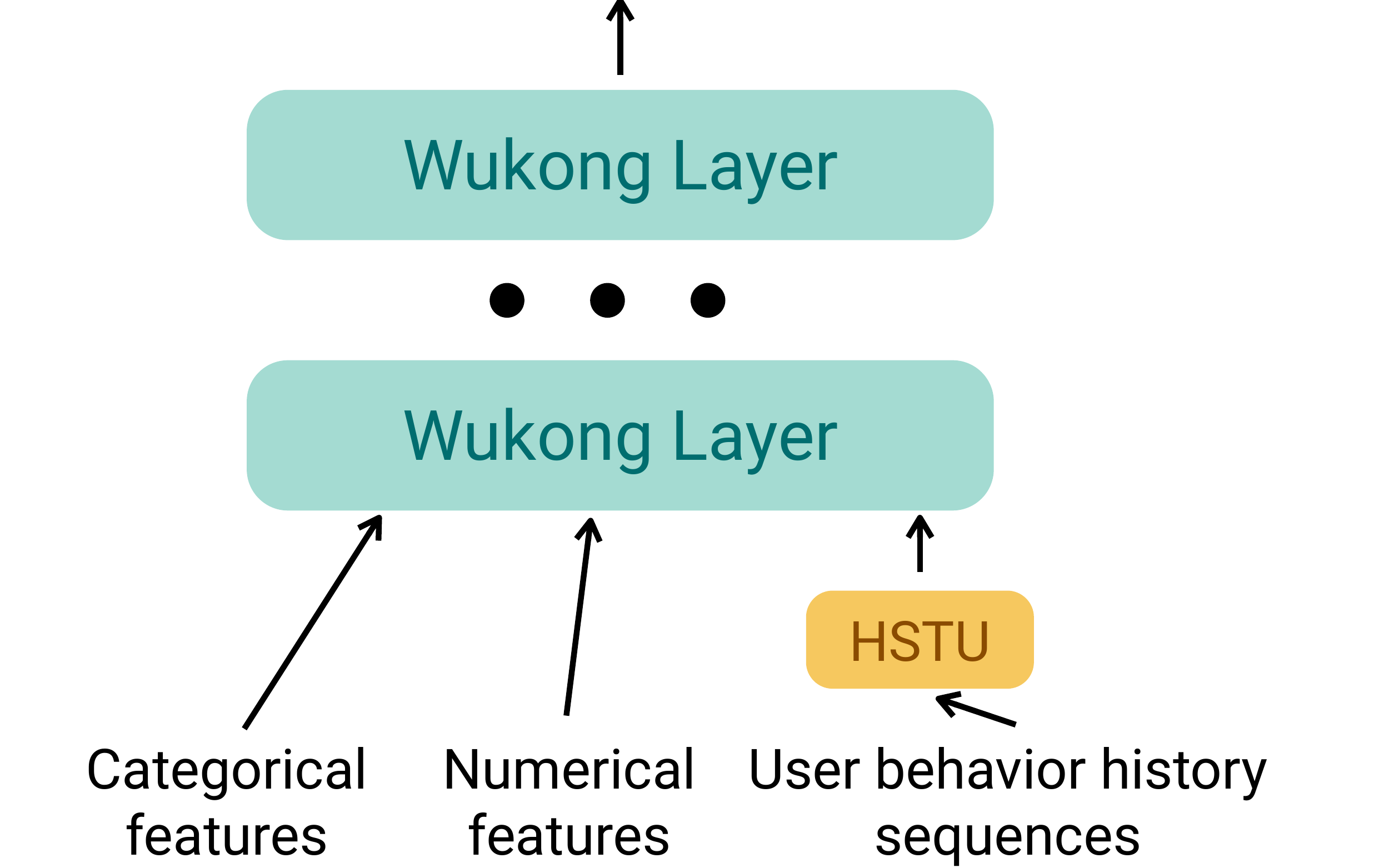}
\caption{Shallow-hybrid architecture.}
\label{fig:hybrid-architecture}
\end{subfigure}\hfill
\begin{subfigure}[t]{0.27\textwidth}
\centering
\includegraphics[width=\linewidth]{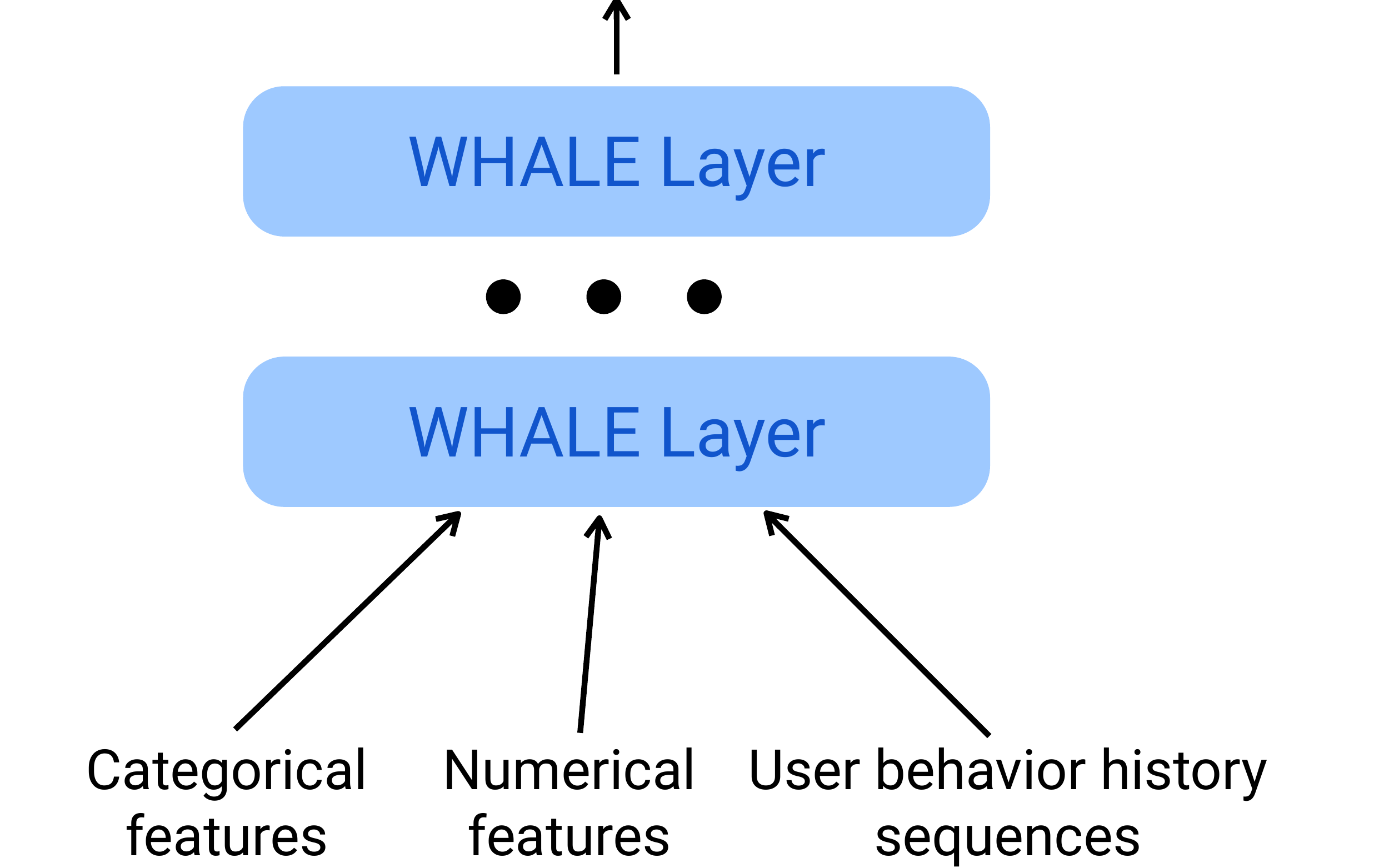}
\caption{\model architecture.}
\label{fig:whale-architecture}
\end{subfigure}
\caption{Overview of Wukong, HSTU, a shallow-hybrid design, and the proposed \model architecture. Wukong provides a scalable backbone for
modeling high-order interactions among non-sequential features, whereas HSTU
provides a scalable backbone for modeling long user-behavior sequences. The
shallow-hybrid design combines these two paradigms by feeding compressed HSTU sequence representations into the Wukong layers. In contrast, \model integrates Wukong and HSTU within every
\model layer, enabling repeated interaction between non-sequential features and sequential behaviors.}
\Description{A row of four architecture diagrams. Subfigure (a) shows the Wukong architecture, subfigure (b) shows the HSTU architecture, subfigure (c) shows the shallow-hybrid architecture, and subfigure (d) shows the WHALE architecture.}
\label{fig:architecture-overview}
\end{figure*}

\section{Introduction}
Large-scale recommender systems are a core component of modern social media platforms~\cite{covington2016youtube,naumov2019dlrm,borisyuk2024lirank,zhu2025rankmixer}. These systems rank candidate user--item interactions at massive scale by combining heterogeneous signals, including long user behavior histories, user and item attributes, contextual features, and explicit user--item crosses, while satisfying strict latency and deployment constraints. As recommendation workloads continue to grow in scale and diversity, increasing model capacity has become an important path toward better quality, because larger models can capture richer user intent, item semantics, and cross-feature patterns~\cite{ardalani2022understanding,zhai2024actions,zhang2024wukong,hou2026kunlun,ha2026unimixer,huang2026hyformer}.

Recent work has advanced large recommender models along two highly scalable but largely separate directions~\cite{zhai2024actions,zhang2024wukong,hou2026kunlun,ha2026unimixer,zhang2025onetrans,zhu2025rankmixer}. One direction focuses on static and contextual feature interactions among non-sequential fields. As illustrated in Figure~\ref{fig:architecture-overview}a, Wukong-style architectures~\cite{zhang2024wukong} demonstrate that recommendation quality can also improve by scaling high-order interactions among user attributes, item attributes, contextual signals, and user--item crosses through factorized parameterizations~\cite{rendle2010factorization} or explicit cross networks~\cite{wang2017deepcross,wang2021dcnv2}. This paradigm is indispensable in ranking systems because many important relevance signals arise from combinations of sparse and dense fields~\cite{gui2023hiformer,guo2017deepfm,lian2018xdeepfm,song2019autoint}. The other direction focuses on dynamic sequential patterns in user behavior histories. As illustrated in Figure~\ref{fig:architecture-overview}b, HSTU-style generative recommenders~\cite{zhai2024actions,zhou2019dien, chai2025longer} show that sequential transduction can effectively model long, heterogeneous interaction sequences and continue to benefit from increased capacity. This paradigm is appealing because user intent evolves over time: recent and historical behaviors reveal changing interests, repeated preferences, and long-range dependencies.

Neither of these two complementary paradigms can fully address the other's core modeling objective on its own. Sequence-centric models can capture temporal evolution in user behavior, but by themselves they do not explicitly scale high-order static and contextual crosses among user, item, and request features. Conversely, feature-interaction models can capture rich non-sequential crosses, but they lack an explicit mechanism for modeling temporal evolution in long behavior sequences. This gap becomes especially clear in social media ranking, where a candidate item's relevance often depends on connecting the current candidate or context to specific events in the user's behavior history. Different candidates or contexts may need to emphasize different parts of the same history: an item--context cross may be supported by a recent action, a repeated long-term preference, or a specific behavioral pattern. Figure~\ref{fig:hybrid-architecture} illustrates a shallow-hybrid way to combine the two paradigms: first use a sequence model to compress the behavior history into a small number of summary embeddings, then feed those summaries together with non-sequential features into a feature-interaction model~\cite{beutel2018latentcross,hou2026kunlun,zhang2025onetrans}. This approach creates a shallow interface between the two signal sources: once the history has been compressed into fixed summaries, high-order feature interactions can no longer selectively draw evidence from fine-grained behavior events as they are formed. This limits progressive exchange between static feature interactions and dynamic sequential patterns.

To address these limitations, we present WHALE (Wukong-HSTU-based scALable unified modEl), a scalable unified recommendation architecture, illustrated in Figure~\ref{fig:whale-architecture}, that keeps Wukong-style feature interaction and HSTU-style sequence modeling as active, capacity-scalable components within a single architecture with progressive interaction. WHALE is built from a stack of recursive layers, where each layer contains a Wukong module, an HSTU module, and an attention-based fusion module. The Wukong module captures high-order interactions among non-sequential features, while the HSTU module models dependencies among behavior-history features. The fusion module then allows the high-order interaction representation from the Wukong branch to query the sequence-aware representation from the HSTU branch, enabling candidate- and context-dependent feature crosses to retrieve the most relevant behavioral evidence from the user's history. By repeating this exchange across layers, WHALE enables progressively deeper exchange between static feature interactions and dynamic sequential patterns.

Our study makes three main contributions. First, we introduce WHALE, a recursive architecture that keeps HSTU and Wukong branches active across layers and fuses them through attention-based cross-branch exchange. To our knowledge, WHALE is the first scalable recommendation architecture that unifies Wukong and HSTU, two representative architectures for non-sequential feature interaction and sequential behavior modeling, within a single model. Second, we develop training and inference optimization techniques that enable the online deployment of WHALE. Third, we show that, through this layer-wise exchange, WHALE scales with sequence length, model depth, and model width, and improves over strong HSTU-only and Wukong-only baselines in both offline evaluation and online A/B tests. Together, these contributions identify progressive Wukong--HSTU cross-branch exchange as a practical design principle for jointly scaling sequential modeling and feature interaction in large industrial recommenders under strict training and serving constraints.
\section{Related work}

\paragraph{Feature interaction models.}
A central line of recommender-system research focuses on modeling
interactions among non-sequential features, including user attributes,
item attributes, contextual signals, and explicit user--item crosses~\cite{beutel2018latentcross,gui2023hiformer}.
Many modern deep ranking models follow an embedding-based pipeline in
which the non-sequential features are first mapped to learned
representations and then combined by an interaction
module~\cite{cheng2016wide,naumov2019dlrm,borisyuk2024lirank}.
Factorization Machines (FM) established an efficient mechanism for
modeling pairwise feature crosses in sparse
settings~\cite{rendle2010factorization}. Subsequent architectures,
including DeepFM~\cite{guo2017deepfm}, xDeepFM~\cite{lian2018xdeepfm},
AutoInt~\cite{song2019autoint}, and DCN-V2~\cite{wang2021dcnv2},
improve the expressiveness of explicit or implicit interaction learning,
enabling ranking models to capture higher-order relationships among
heterogeneous fields.

\noindent\textbf{Wukong.} Within this family, Wukong is one of the most representative work, which scales the interaction module itself~\cite{zhang2024wukong}. Its core design is a stacked
factorization-machine architecture built from a factorization machine
block (FMB) and a linear compression block (LCB). Let $X_i$ denote the
feature-embedding representation entering the $i$-th Wukong interaction
layer, and let $X_{i+1}$ denote the transformed representation passed to
the next layer. A Wukong interaction layer can then be summarized as
\begin{equation}
\label{eq:wukong-layer}
\begin{aligned}
\operatorname{FMB}_i(X_i)
&= \operatorname{reshape}\!\left(
\operatorname{MLP}\!\left(
\operatorname{LN}\!\left(
\operatorname{flatten}\!\left(\operatorname{FM}(X_i)\right)
\right)\right)\right), \\
\operatorname{LCB}_i(X_i)
&= W_L X_i, \\
X_{i+1}
&= \operatorname{LN}\!\left(
\operatorname{concat}\!\left(\operatorname{FMB}_i(X_i), \operatorname{LCB}_i(X_i)\right) + X_i
\right).
\end{aligned}
\end{equation}
Here, $\operatorname{FM}(\cdot)$ denotes the FM
interaction operator, $\operatorname{MLP}(\cdot)$ denotes a multilayer
perceptron, and $\operatorname{LN}(\cdot)$ denotes layer normalization,
which stabilizes feature representations across the recursive
interaction layers; $W_L$ denotes the learnable linear projection in the
LCB. The FMB explicitly models pairwise interactions and maps them back
to embedding space, whereas the LCB linearly recombines the input
features. By recursively repeating
this process, deeper stacks can represent progressively higher-order
feature crosses. This makes depth and width natural
scaling knobs for large-scale ranking systems. In our setting, Wukong
provides a strong and scalable backbone for non-sequential user, item,
contextual, and cross features, but by itself
it does not model the temporal evolution and long-range dependencies
encoded in user behavior sequences.

\paragraph{Sequence-aware recommendation.}
Another important line studies user behavior histories as ordered
event sequences. Unlike
non-sequential feature-interaction models, sequence-aware recommenders
preserve temporal order and infer the user's current intent from
recency patterns, repeated interests, and long-range dependencies in
historical behavior. DIN and DIEN showed
that item-conditioned attention and explicit interest evolution over
behavior sequences can significantly improve ranking
quality~\cite{zhou2018din,zhou2019dien}. More recently, LONGER studies
the scaling of ultra-long behavior-sequence modeling in industrial
recommenders~\cite{chai2025longer}. Together, these works show that
temporal context and evolving interests are central recommendation
signals that should be modeled directly from user logs rather than only
through static feature crosses.

\noindent\textbf{HSTU.} Within this line, HSTU is one of the most representative works~\cite{zhai2024actions,ding2026bending}, which is designed for long, and non-stationary behavior sequences. It captures long-range user dynamics while continuing to benefit from increased model and training compute. Borrowing the notation of the original paper, let $X_i$ denote the input to the $i$-th HSTU layer. A compact HSTU block can then be summarized as
\begingroup
\setlength{\abovedisplayskip}{3pt}
\setlength{\belowdisplayskip}{3pt}
\setlength{\abovedisplayshortskip}{2pt}
\setlength{\belowdisplayshortskip}{2pt}
\begin{equation}
\label{eq:hstu-layer}
\begin{aligned}
\left[U_i, V_i, Q_i, K_i\right]
&= \operatorname{Split}\!\left(\phi\!\left(\operatorname{Norm}(X_i)W_1 + b_1\right)\right), \\
A_i
&= \operatorname{Norm}\!\left(\phi\!\left(Q_iK_i^{\top} + \mathrm{rab}^{(p,t)}\right)V_i\right), \\
X_{i+1}
&= X_i + W_2\!\left(A_i \odot U_i\right) + b_2,
\end{aligned}
\end{equation}
\endgroup
where $W_1$ and $b_1$ denote the learnable projection matrix and bias used to produce $U_i$, $V_i$, $Q_i$, and $K_i$; $W_2$ and $b_2$ denote the learnable output projection matrix and bias; $\phi$ denotes the pointwise nonlinearity (e.g., SwiGLU); and $\mathrm{rab}^{(p,t)}$ denotes the relative attention bias over positional and temporal information. As a scaling backbone for sequence-aware recommendation, HSTU captures dynamic user intent, but by itself it does not explicitly capture the rich high-order interactions among non-sequential user, item, and cross features, which motivates us combining it with Wukong.

\paragraph{Hybrid recommendation architectures.}
More recent work moves toward joint modeling of sequential and non-sequential
signals. UniMixer and RankMixer are closer to efforts that improve the
scalability of dense ranking backbones: UniMixer unifies attention-based,
TokenMixer-based, and FM-based scaling blocks within a
common feature-mixing framework~\cite{ha2026unimixer}, while RankMixer proposes
a hardware-aware token-mixing architecture for scalable industrial
ranking~\cite{zhu2025rankmixer}. These models improve scalable feature mixing,
but they typically incorporate sequential information through compact summaries
rather than maintaining a fine-grained history branch that repeatedly interacts
with non-sequential features. Such shallow coupling limits progressive
cross-branch refinement as model depth grows.

Kunlun is more closely related, as it proposes a multi-layer architecture in
which each layer contains a sequence modeling block and an interaction block for
bidirectional exchange between sequence and non-sequence
features~\cite{hou2026kunlun}. This is an important step toward unified
recommendation modeling, but it differs from our setting: Kunlun does not adopt
an HSTU-style sequence backbone and therefore does not directly target
request-level reuse of user-history computation~\cite{guo2026request}. This reuse is especially
important in large-traffic organic ranking, e.g., short-form video
recommendation, where many candidates in the same request share the same user
history; computing the history representation once and reusing it across
candidates makes long-sequence modeling computationally feasible for large-scale
industrial serving. OneTrans provides another unification
approach~\cite{zhang2025onetrans}, but it is not built around a separately
scalable feature-interaction backbone specialized for non-sequential feature
interactions, as Wukong is. \model instead explicitly builds on Wukong and HSTU, two scalable designs that
have been validated in large-scale industrial recommendation
settings~\cite{zhang2024wukong,zhai2024actions}. By keeping both backbones
active inside each recursive layer, \model allows high-order non-sequential
interactions to repeatedly query fine-grained behavior-history signals rather
than relying only on compact sequence summaries or non-sequential interactions
alone.

\paragraph{Scaling laws for recommendation.}
Recent work has begun to ask whether recommendation models obey scaling laws similar to those observed in language models. Existing studies on DLRM-style recommenders characterized how quality varies with parameters, data, and compute~\cite{ardalani2022understanding}, while HSTU and Wukong push this discussion further from two complementary directions: HSTU emphasizes large-scale sequential transduction~\cite{zhai2024actions}, whereas Wukong demonstrates that dense feature-interaction modules can also exhibit favorable scaling behavior~\cite{zhang2024wukong}. Our work builds on these observations and studies how a unified architecture can benefit from both sequential and feature-interaction branches in a practical industrial organic content recommendation
setting.
\section{Model architecture design}

\subsection{Overview}
As shown in Fig.~\ref{fig:whale-layer-architecture}, the proposed model consists of an input layer followed by stacked WHALE layers. The input layer maps heterogeneous raw features into embeddings. Each WHALE layer then contains a Wukong module for high-order non-sequence feature interaction, an HSTU module for sequence modeling, and an attention-based fusion module for progressive cross-branch exchange between the Wukong module and the HSTU module. In the remainder of this section, we first describe the input layer and then the internal structure of a WHALE layer.

\subsection{Input layer}
\model takes in diverse input signals, including user and item categorical features, user and item numerical features, and user behavior sequences. The input layer maps these heterogeneous features into dense embeddings, which serve as the initial representations for the stacked WHALE layers.

\begin{figure}[t]
\centering
\includegraphics[width=0.9\columnwidth,trim=0 4 0 0,clip]{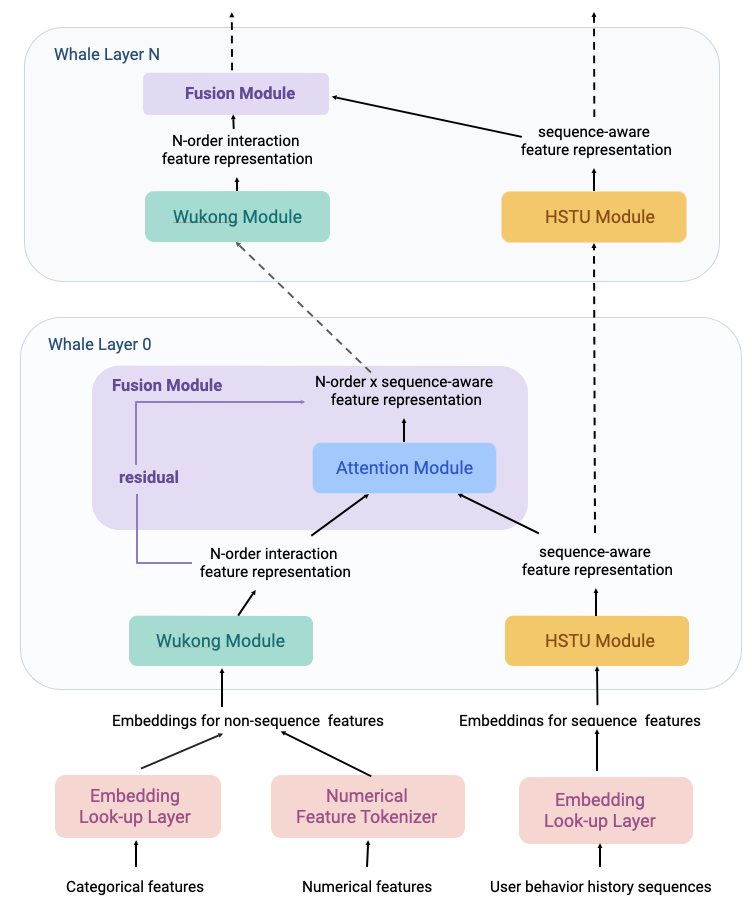}
\vspace{-0.2em}
\caption{Detailed overview of the \model{} architecture. The input layer maps
raw categorical, numerical, and sequence features into dense embeddings
using embedding lookup tables and a numerical feature tokenizer. Stacked
WHALE layers then combine Wukong, HSTU, and attention-based fusion modules
to produce joint feature representations, which are used to generate task
predictions.}
\Description{A WHALE architecture diagram with feature embedding modules at the bottom, a detailed WHALE layer 0 in the middle containing Wukong, HSTU, attention, fusion, and residual connections, and repeated WHALE layers above it.}
\label{fig:whale-layer-architecture}
\end{figure}

For non-sequence features, categorical fields such as user IDs or item IDs are mapped to learnable embeddings through embedding tables. Numerical fields,such as intensity or count features, are mapped to embeddings using MLPs; this MLP-based component is labeled as the ``Numerical Feature Tokenizer'' in Figure~\ref{fig:whale-layer-architecture}. We then fuse the categorical and numerical embeddings into non-sequence feature representations, denoted as $E_{ns} \in \mathbb{R}^{N \times D}$, where $N$ is the number of non-sequence
features after the input layer.

For sequence features, the user behavior history records the items that a user interacted with before the current request. Each historical behavior is represented by an item ID together with side information, such as action type and author ID. We first map these discrete attributes into embeddings using embedding layers, and then use MLPs to fuse the item and side-information embeddings into sequence feature representations, denoted as $E_s \in \mathbb{R}^{L \times D}$, where $L$ is the sequence length. Although the sequence and non-sequence representations need not share the same dimensionality, we assume a common embedding dimension $D$ for notational simplicity.

\subsection{\layer}
Each WHALE layer is designed to let candidate- and context-dependent non-sequence feature crosses retrieve relevant evidence from the user's behavior history. Specifically, for the input non-sequence representations and sequence representations at layer $\ell$, the Wukong module constructs high-order interaction representations over user, item, and contextual features, while the HSTU module encodes fine-grained temporal evidence from the behavior history. The fusion module then uses the Wukong output to query the HSTU output, injecting the most relevant sequential signals into the non-sequence interaction representation. By stacking multiple WHALE layers, \model enables progressive cross-branch exchange, so static feature interactions and dynamic sequence patterns can repeatedly condition on one another.

We initialize the layer-wise non-sequence and sequence representations as $X_{ns}^{(0)} = E_{ns}$ and $X_s^{(0)} = E_s$, respectively. At the $\ell$-th WHALE layer, $X_{ns}^{(\ell)} \in \mathbb{R}^{N \times D}$ and $X_s^{(\ell)} \in \mathbb{R}^{L \times D}$ denote the current non-sequence and sequence representations.

\subsubsection{Wukong module: interaction of non-sequence features}
The Wukong module operates on the non-sequence branch, which contains user-profile features, item-attribute features, contextual features, and explicit user--item cross features. Although these features do not have a temporal order, their higher-order combinations are highly informative for ranking. For example, a candidate item's relevance may depend not only on the user's interests or the item's attributes individually, but also on how the candidate, user, and request context interact. The Wukong module therefore constructs candidate- and context-dependent interaction representations that capture these ranking signals. Additionally, these feature representations serve as queries for retrieving relevant behavioral evidence from the sequence branch.

Following the Wukong module reviewed in Related Work (Eq.~\eqref{eq:wukong-layer}), it applies stacked feature-interaction transformations to $X_{ns}^{(\ell)}$ and produces an $n$-order interaction representation,
\[
H_{w}^{(\ell)} = \text{Wukong\_module}(X_{ns}^{(\ell)}) \in \mathbb{R}^{N \times D}.
\]

\subsubsection{HSTU module: modeling behavior sequences}
Complementary to the Wukong branch, the HSTU module operates on the sequence branch and focuses on modeling the temporal evolution of user interests. Its input is the ordered behavior history, where each step is represented by a unified feature representation constructed from the interacted item and its associated side information, e.g., author information and timestamp information. Compared with non-sequence features, these representations carry both semantic content and chronological structure, so the module captures recency effects, repeated interests, and long-range dependencies across behavior steps~\cite{zhai2024actions}. The HSTU module therefore contextualizes each behavior step with the rest of the history, preserving fine-grained behavioral evidence that can later be used by the fusion module.

Following the HSTU block reviewed in Related Work (Eq.~\eqref{eq:hstu-layer}), it applies stacked sequential transformations to $X_{s}^{(\ell)}$ and produces sequence-aware behavior representations,
\begin{equation}
H_{s}^{(\ell)} = \text{HSTU\_module}(X_{s}^{(\ell)}) \in \mathbb{R}^{L \times D}.
\label{eq:hstu-module-output}
\end{equation}
The output $H_{s}^{(\ell)}$ remains an $L$-step representation of the
behavior history. This granularity is important for \model: different
Wukong interaction representations may need different behavioral evidence,
such as recent actions, or a specific
pattern involving action type, author information, and temporal signals
such as the time gap between the action and the current request.

\subsubsection{Fusion module: attention-based cross-branch fusion}
The fusion module provides the main cross-branch exchange mechanism in \model:
it connects high-order non-sequence interaction representations with
fine-grained behavior-history representations. Rather than simply
concatenating the outputs of the Wukong and HSTU branches, it uses
cross-branch attention~\cite{vaswani2017attention}, where Wukong interaction representations act as
queries and HSTU behavior representations provide keys and values. This
allows each candidate- and context-dependent interaction representation to
selectively retrieve relevant behavioral evidence from the user's history.

Concretely, let $H_{w}^{(\ell)} \in \mathbb{R}^{N \times D}$ and
$H_{s}^{(\ell)} \in \mathbb{R}^{L \times D}$ denote the outputs of the
Wukong and HSTU modules at layer $\ell$, respectively. With pre-layer
normalization, the attention projections are computed as
\[
\widehat{H}_{w}^{(\ell)} = \mathrm{LN}_{w}^{(\ell)}(H_{w}^{(\ell)}), \qquad
\widehat{H}_{s}^{(\ell)} = \mathrm{LN}_{s}^{(\ell)}(H_{s}^{(\ell)}),
\]
\[
Q^{(\ell)} = \widehat{H}_{w}^{(\ell)} W_Q^{(\ell)}, \qquad
K^{(\ell)} = \widehat{H}_{s}^{(\ell)} W_K^{(\ell)}, \qquad
V^{(\ell)} = \widehat{H}_{s}^{(\ell)} W_V^{(\ell)},
\]
where $\mathrm{LN}_{w}^{(\ell)}$ and $\mathrm{LN}_{s}^{(\ell)}$ are layer
normalization operations, and $W_Q^{(\ell)}, W_K^{(\ell)}, W_V^{(\ell)} \in
\mathbb{R}^{D \times D}$ are learnable projection matrices. The attention
output is then given by
\begin{equation}
A^{(\ell)} = \mathrm{softmax}\left(\frac{Q^{(\ell)} {K^{(\ell)}}^{\top}}{\sqrt{D}}\right)V^{(\ell)} \in \mathbb{R}^{N \times D}.
\label{eq:cross-attention}
\end{equation}

The resulting $A^{(\ell)}$ organizes the attended behavior information
into $N$ interaction-specific representations, one for each Wukong
interaction representation. Since the attention weights are
computed between Wukong-derived queries and HSTU-derived keys, each row of
$A^{(\ell)}$ aggregates behavior-history evidence that is relevant to a
particular candidate- and context-dependent non-sequence interaction.

To further fuse the two branches, we introduce a \emph{Fusion MLP} that
combines the attention output with the corresponding Wukong representations.
Specifically, the Fusion MLP concatenates $A^{(\ell)}$ with
$H_{w}^{(\ell)}$, projects the concatenated representation back to dimension
$D$, and applies a residual connection:
\begin{equation}
\overline{A}^{(\ell)} = H_{w}^{(\ell)} + [H_{w}^{(\ell)} | A^{(\ell)}] W_F^{(\ell)}
\in \mathbb{R}^{N \times D},
\label{eq:fusion-concat-projection}
\end{equation}
where $W_F^{(\ell)} \in \mathbb{R}^{2D \times D}$ is the learnable projection
matrix of the Fusion MLP, and $[\cdot | \cdot]$ denotes concatenation along the
feature dimension. Intuitively, $\overline{A}^{(\ell)}$ combines candidate- and
context-conditioned behavioral evidence with the high-order interaction
features.

Following the standard attention-block
design~\cite{vaswani2017attention}, we further apply a pre-norm feed-forward
network (FFN) to transform this fused representation. We implement the FFN as a
SwiGLU FFN, which uses a gating mechanism to modulate one projected
representation by another and has shown improved modeling
quality~\cite{shazeer2020glu}. Applied to $\overline{A}^{(\ell)}$, it is
computed as
\begin{equation}
\begin{aligned}
\widehat{A}^{(\ell)}
&= \mathrm{LN}_{\mathrm{ffn}}^{(\ell)}(\overline{A}^{(\ell)}), \\
\widetilde{A}^{(\ell)}
&= \overline{A}^{(\ell)} + \mathrm{SwiGLU\_FFN}^{(\ell)}(\widehat{A}^{(\ell)}) \\
&= \overline{A}^{(\ell)} + \bigl(\mathrm{SiLU}(\widehat{A}^{(\ell)} W_{\mathrm{gate}}^{(\ell)}) \odot
\widehat{A}^{(\ell)} W_{\mathrm{up}}^{(\ell)}\bigr) W_{\mathrm{down}}^{(\ell)}
\in \mathbb{R}^{N \times D},
\end{aligned}
\label{eq:swiglu-ffn}
\end{equation}
where $\mathrm{LN}_{\mathrm{ffn}}^{(\ell)}$ is a layer normalization operation,
$\mathrm{SiLU}(\cdot)$ denotes the SiLU activation, $\odot$ denotes
element-wise multiplication, and $W_{\mathrm{gate}}^{(\ell)}$,
$W_{\mathrm{up}}^{(\ell)}$, and $W_{\mathrm{down}}^{(\ell)}$ are learnable
projection matrices. The resulting $\widetilde{A}^{(\ell)}$ is therefore a
residual gated refinement of the fused cross-branch representation.

The output of the pre-norm SwiGLU FFN is used as the fused representation and
passed to the next WHALE layer:
\begin{equation}
X_{ns}^{(\ell+1)} = H_{f}^{(\ell)} = \widetilde{A}^{(\ell)}.
\label{eq:fusion-output}
\end{equation}
The resulting fused representation $H_{f}^{(\ell)}$ is thus both sequence-aware
and interaction-aware.
In parallel, the HSTU output from Eq.~\eqref{eq:hstu-module-output} is
propagated along the sequence branch and serves as the sequence input to the
HSTU module in the next layer:
\[
X_s^{(\ell+1)} = H_s^{(\ell)}.
\]
In this way, \model preserves separate propagation paths for both
non-sequence and sequence representations while allowing them to exchange
information through the fusion module at each layer.
\section{Efficiency optimization}
\label{sec:efficiency-optimization}

Recommendation system training and inference are among the most demanding
GPU workloads, involving high-throughput data processing, latency-sensitive
serving, and irregular tensor shapes that depend on input features.
Consequently, improving system efficiency is necessary to reduce training
cost and satisfy the online serving latency and throughput constraints.

In this section, we describe the training and inference efficiency
optimizations that enable WHALE's online deployment. We use queries per
second (QPS) as the throughput metric. Training QPS denotes the number
of samples processed per second by the distributed training system, while
inference QPS denotes the number of requests served per second under the
target latency budget.

\subsection{Training QPS optimization}
Training QPS optimization builds on the optimized implementations of the
Wukong and HSTU modules and addresses the additional efficiency bottlenecks
introduced by \model. We combine customized kernels, lightweight
architectural changes, mixed-precision execution, and compiler-level graph
optimization to increase throughput while preserving model quality.

\paragraph{Customized attention kernel.}
We implement cross-branch attention in the fusion module with a fused
Triton kernel~\cite{tillet2019triton}. For \model, the kernel incorporates
the following customizations:

\noindent\emph{(1) Shared key--value.} Profiling shows that cross-branch attention
kernels are memory-bound. To reduce this overhead, we
share the key and value tensors in \model, i.e., $K{=}V$. The fused kernel can then reuse
the loaded key tensor as the corresponding value tensor,
avoiding separate value-tensor reads and halving KV-side
GPU memory traffic. In the backward pass, the shared KV gradient is accumulated
inside the kernel, further reducing memory traffic and improving kernel
throughput. Empirically, we observe this change does not degrade model quality.

\noindent\emph{(2) Asymmetric backward variants.} In \model, cross-branch
attention has an asymmetric shape: a small number of query tokens (e.g., 64)
attend to a long user-history sequence (e.g., 15,000). This asymmetric shape
makes the backward pass sensitive to how the computation is partitioned,
because query-parallel and key--value-parallel partitioning incur different
gradient-reduction costs.

We use the FlashAttention backward schedule~\cite{dao2024flashattention2}
as the \emph{KV-parallel} variant, which partitions the backward computation
along the key--value dimension, accumulates key--value-side gradients locally,
and reduces query-side gradients across workers. For short-query
regime, we add a complementary \emph{Q-parallel} variant that partitions the
backward computation along the query dimension, allowing query-side gradients
to remain local while scanning the long history sequence. We select between
the two variants at runtime based on the query and history lengths. This
shape-aware selection reduces unnecessary memory traffic and yields about
1.2x kernel-level speedup over using only the KV-parallel baseline.
  
\paragraph{Shared-gate SwiGLU FFN} As described in Eq.~\ref{eq:swiglu-ffn}, the
SwiGLU~\cite{shazeer2020glu} FFN uses separate gate, up, and down
projection matrices, requiring three matrix multiplications. To improve QPS,
we use a shared-gate variant by tying the gate and up projection matrices,
$W_{\mathrm{gate}}^{(\ell)} = W_{\mathrm{up}}^{(\ell)}
\triangleq W_s^{(\ell)}$, yielding
\begin{equation}
\begin{aligned}
\mathrm{SharedGateSwiGLU\_FFN}^{(\ell)}(\overline{A}^{(\ell)})
&= \left(
\mathrm{SiLU}\!\left(\overline{A}^{(\ell)} W_s^{(\ell)}\right)
\odot \overline{A}^{(\ell)} W_s^{(\ell)}
\right) \\
&\quad W_{\mathrm{down}}^{(\ell)}.
\end{aligned}
\end{equation}
This computes the shared projection only once, reducing the number of FFN
matrix multiplications from three to two, i.e., a 33\% reduction.
Empirically, we observe this weight tying does not degrade model quality.

\paragraph{Mixed-precision training and compiler optimization.}
We adopt two widely used efficiency techniques---mixed-precision execution and
compiler-level graph optimization~\cite{gruslys2016memory, micikevicius2018mixed,kaplan2020scaling,zhang2025onetrans}---and apply them selectively according to the computational
characteristics of \model.
For mixed-precision execution, the dense backbone---including Wukong modules,
HSTU modules, fusion modules, and all FFNs---runs in BF16 to reduce memory bandwidth
and improve throughput, whereas numerically sensitive components such as
loss computation, and task arch remain in FP32 to
preserve training stability~\cite{micikevicius2018mixed}. For compiler-level
graph optimization, we apply \texttt{torch.compile} to the dense backbone
subgraphs, enabling TorchInductor to fuse operators, plan memory usage, and
dispatch compiled kernels more efficiently~\cite{gruslys2016memory}.

\subsection{Inference QPS optimization}

Several training-side efficiency optimizations also improve inference QPS, including the
shared-gate SwiGLU FFN, customized attention kernels, and compiler-level
graph optimization. However, online serving introduces additional bottlenecks
that are less visible during training, such as irregular tensor shapes, kernel
launch overhead, and CPU--GPU synchronization caused by runtime shape
computation. We therefore further optimize the inference path through
shape-aware kernel tuning, compiled graph execution, and synchronization-free
shape handling.
\paragraph{Kernel-level optimization \& operator fusion.}
On top of the Triton kernels customized to optimize training efficiency, we further customize the Triton kernels to optimize inference efficiency. Specifically, we tune the shape of the tensors in the kernel, regarding to the performance on serving traffic. To further streamline performance, we enabled AOTInductor\cite{aotinductor}, which extends PyTorch’s compiler stack by ahead-of-time compiling, to group operations and merge each group of operations into a single fused operation. This approach reduces memory access frequency and minimizes kernel launch overhead by combining multiple operations into single, efficient passes.
\cite{goto2008anatomy,dao2022flashattention,nvidia2026matmulguide}.
This optimization yields 18\% higher throughput than irregular-shape
alternatives.

\paragraph{Whole-graph optimization for dynamic shapes.}
In models involving dynamic tensor shapes, frequent GPU-CPU synchronization often creates significant bottlenecks during runtime shape calculation. We addressed this by implementing shape hint tensors, which allow the system to infer dimensions without triggering costly synchronizations, ensuring a more fluid, GPU-centric execution flow.
\cite{pytorch2025nonzero,pytorch2025syncdebug}.
Eliminating a single synchronization point can reduce serving latency by
\mbox{$1$--$5\,\mathrm{ms}$}, which is critical for serving \model as a large ranking
model under tight latency constraints. This optimization increases inference
QPS by 15\%.

\section{Experiments}

We evaluate \model in a large-scale industrial recommendation setting and through both offline and online experiments. In this section, we aim to answer the following three questions:

\textbf{Q1: How well does \model scale?}
We study the scaling behavior of \model through two sets of experiments. First,
we compare \model with state-of-the-art baselines across different model scales
to evaluate whether \model consistently achieves better quality as model scale
increases.
Second, we vary the sequence length of the user interaction history, model
depth, and model width within \model to test whether the proposed architecture
can effectively benefit from more historical context and larger model capacity.

\textbf{Q2: How important is the fusion module?}
We ablate key components of the fusion module to understand how cross-branch
interaction contributes to recommendation quality, e.g.,the attention, the fusion MLP, and the SwiGLU FFN.

\textbf{Q3: How does \model perform online?}
We deploy \model in an online A/B test under the same serving setup as the online baseline. This experiment verifies
whether \model delivers measurable gains in real recommendation traffic.

\subsection{Offline experiment settings}
For the offline study, we use training logs from one of the largest short-form
video social media platforms, with 80B samples for training and 4B samples for
evaluation. Unless otherwise stated, all experiments are conducted on NVIDIA B200 GPUs, with the global batch size and optimizer settings held constant across runs. Each model is trained for a single epoch. We report normalized entropy (NE)~\cite{he2014practical} for the platform's
primary engagement prediction task. Using binary labels $y_i \in \{0,1\}$, predicted probabilities $p_i$, and empirical click-through rate $\bar{p}$,  NE is defined as
\begin{equation}
\mathrm{NE} = \frac{-\frac{1}{N} \sum_{i=1}^{N} \left[y_i \log p_i + (1-y_i) \log(1-p_i)\right]}{-\left(\bar{p} \log \bar{p} + (1-\bar{p}) \log(1-\bar{p})\right)}.
\end{equation}
Lower NE indicates better model quality. On this platform, a 0.05\% NE gain is
considered a noticeable win. Unless otherwise stated, FLOPs denote per-example
forward-pass computation and are used as a relative model-complexity measure
across architectures and scaling configurations.\footnote{Note that the FLOPs reported in this paper include request-only computations. For example, the K/V projection in \model needs to be performed only once per request, even if the request contains hundreds of examples (i.e., candidates). In practice, we use the M-FALCON algorithm in~\cite{zhai2024actions} to amortize these computational costs.}

\subsection{Architecture comparison}
\begin{figure}[htbp]
\centering
\includegraphics[width=0.8\columnwidth,trim=7 9 6 7,clip]{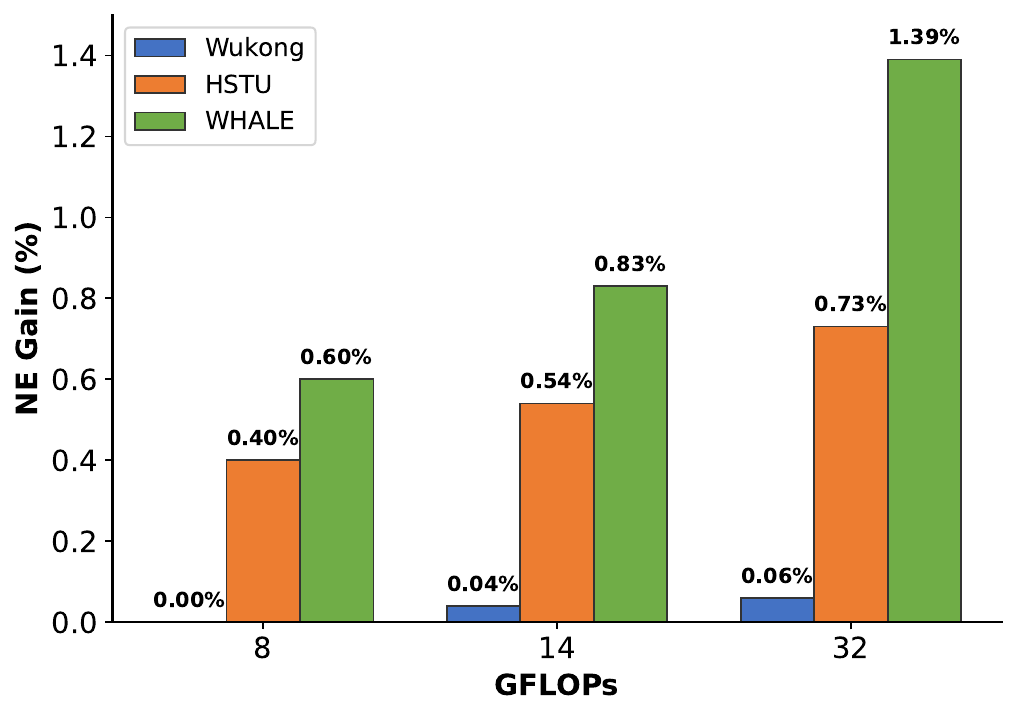}
\vspace{-0.6em}
\caption{Model-quality comparison between \model and single-paradigm baselines at
8, 14, and 32 GFLOPs. The 8-GFLOP Wukong-only model serves as the
reference baseline, and the y-axis reports NE gains relative to this baseline.
Larger NE gains indicate better model quality.}
\Description{A bar chart comparing the proposed model with Wukong-only and HSTU-only baselines.}
\label{fig:architecture-comparison}
\end{figure}

We compare \model against two strong single-paradigm baselines that have been
validated through online experiments:
\begin{itemize}
\item \textbf{Wukong-only ranker}, which emphasizes high-order interactions
among non-sequence features.
\item \textbf{HSTU-only ranker}, which focuses on sequence modeling over user
behavior histories.
\end{itemize}
Comparing against these two baselines helps quantify the utility of unifying
feature-interaction modeling and sequence modeling in \model. To isolate the
benefit of this unified design, we align the model complexity of the three
models as closely as possible by adjusting their hidden dimensions, number of
layers and sequence length, and we use FLOPs to measure model complexity. As shown in
Fig.~\ref{fig:architecture-comparison}, \model consistently outperforms the two
single-paradigm baselines as model complexity increases across the evaluated
range. This result suggests that unifying the two paradigms improves
scalability: rather than increasing the capacity of either paradigm alone,
\model benefits from complementary signals from both feature interactions and
behavior sequences, enabling better model quality under comparable complexity. 

\subsection{Scaling analysis}
\begin{figure*}[htbp]
\centering
\begin{subfigure}[t]{0.30\textwidth}
\centering
\includegraphics[width=0.9\linewidth,trim=7 7 8 6,clip]{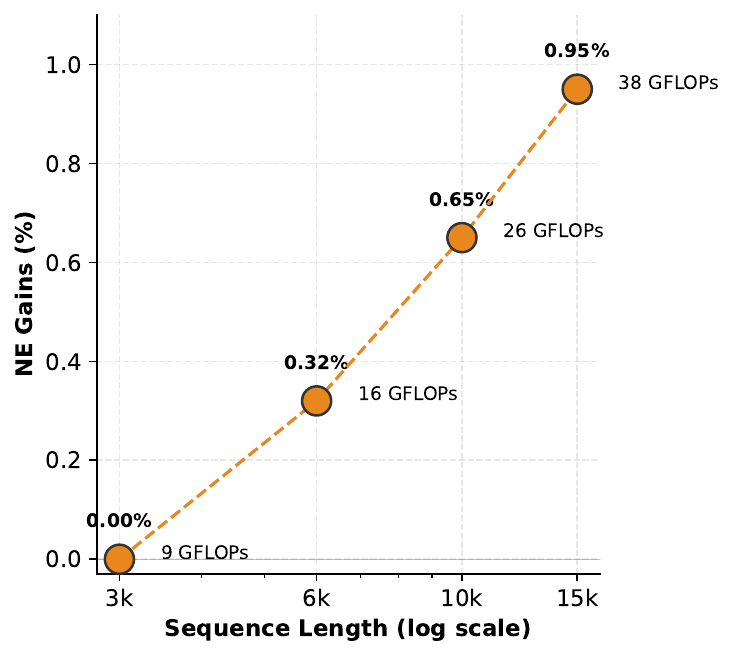}
\caption{Sequence length scaling.}
\label{fig:sequence_scaling}
\end{subfigure}\hfill
\begin{subfigure}[t]{0.30\textwidth}
\centering
\includegraphics[width=0.9\linewidth,trim=7 7 8 6,clip]{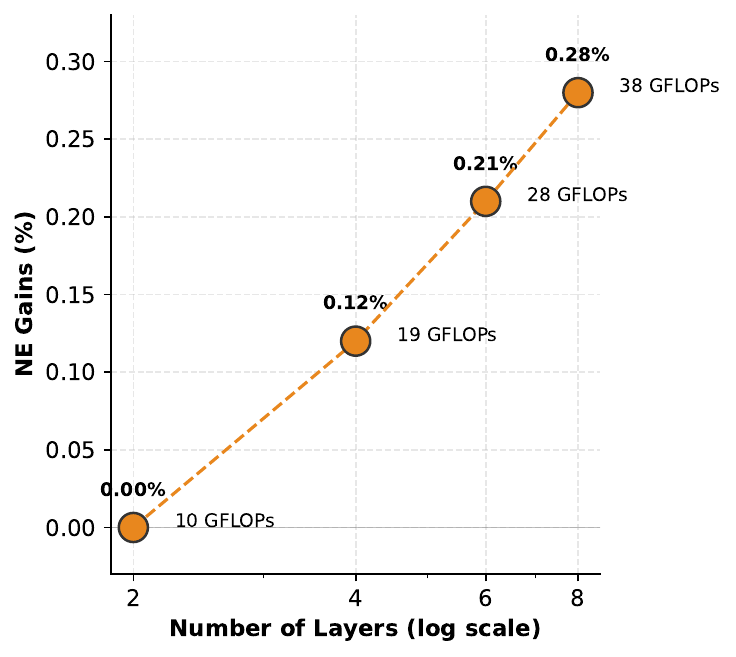}
\caption{Model depth scaling.}
\label{fig:depth_scaling}
\end{subfigure}\hfill
\begin{subfigure}[t]{0.30\textwidth}
\centering
\includegraphics[width=0.9\linewidth,trim=7 7 8 6,clip]{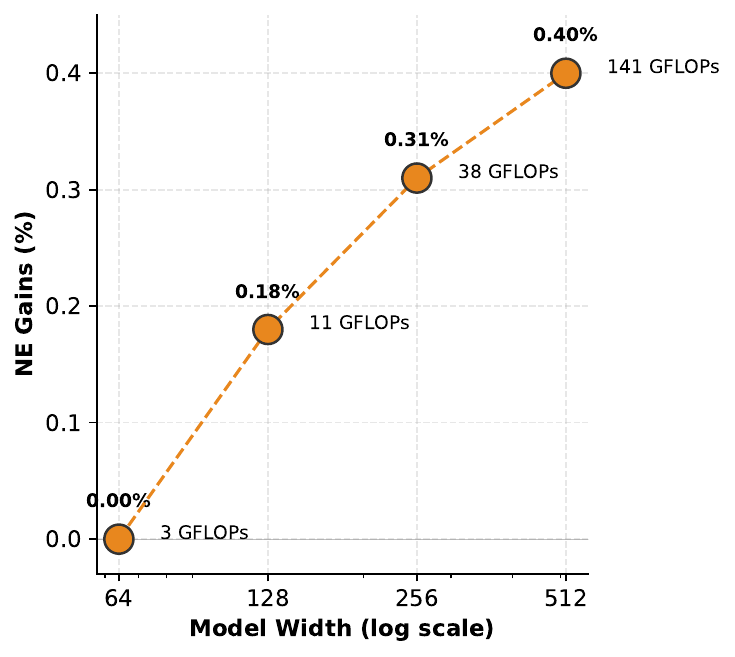}
\caption{Model width scaling.}
\label{fig:width_scaling}
\end{subfigure}
\caption{Scaling analysis of \model with respect to sequence length, model depth, and model width. In the sequence length scaling study, the model with 3k sequence features serves as the baseline, and NE gains for the 6k, 10k, and 15k settings are reported relative to it. In the depth scaling study, the 2-layer model is the baseline; in the width scaling study, the model with a 64-dimensional embedding is the baseline. The annotations on each curve denote the corresponding NE gains and FLOPs.}
\Description{A three-panel figure showing scaling plots for the proposed model with respect to input user interaction history length, model depth, and model width.}
\label{fig:scaling_law_overview}
\end{figure*}

In this subsection,
we analyze the scaling behavior of \model with respect to models' computation complexity by varying the sequence length, model width, and model depth. 

\paragraph{\textbf{Sequence length scaling}}
We study how \model scales with the sequence length of the user interaction history by
varying the sequence length among 3k, 6k, 10k, and 15k while keeping all other
configurations fixed. Specifically, we use an 8-layer \model and set
the embedding dimensions of both non-sequence and sequence features to 256. The
results in Fig.~\ref{fig:sequence_scaling} show that increasing the sequence
length consistently produces NE gains up to 15k, the longest history length
available in our dataset. This trend indicates that \model can effectively exploit longer
behavior histories and exhibits favorable scaling behavior with respect to
sequence length.

\paragraph{\textbf{Model depth scaling}}
We then evaluate the effect of model depth by increasing the number of \model
layers from 2 to 4, 6, and 8. For this study, we keep the sequence length fixed
at 15k and use 256-dimensional embeddings for both non-sequence and sequence
features. As shown in Fig.~\ref{fig:depth_scaling}, deeper models consistently
produce NE gains up to 8 layers. This result suggests that stacking additional
\model layers strengthens the progressive exchange between the feature
interaction and sequence branches, leading to better scaling with model depth.

\paragraph{\textbf{Model width scaling}}
To evaluate the impact of model width capacity, we vary the embedding
dimension shared by the non-sequence and sequence branches from 64 to 128, 256, and 512. The sequence length and depth are fixed to 15k and 8 layers,
respectively. Fig.~\ref{fig:width_scaling} shows that increasing the width
continues to yield NE gains. These results indicate
that \model can make effective use of wider branch representations, further
supporting its scalability with respect to model capacity.

Overall, the three scaling studies show that \model consistently benefits from
more behavioral context, deeper cross-branch computation, and wider
representations. These results indicate that \model exhibits consistent scaling
behavior across multiple sources of model capacity.

\subsection{Study of different fusion strategies}
The fusion strategy determines how \model exchanges information between
non-sequence feature interactions and user behavior sequences. We study this
design choice from two levels, i.e., the architecture level and the module level, as summarized in
Table~\ref{tab:fusion-strategy-study}. At the architecture level, we compare
the proposed progressive fusion in \model with the shallow-hybrid variant
illustrated in Fig.~\ref{fig:architecture-overview}, which uses a shallower
fusion design. At the module level, we focus only on the fusion module in
\model and study its internal design choices by replacing the attention
operation in Eq.~\eqref{eq:cross-attention} with average pooling over the
sequence-length dimension of $V^{(\ell)}$, removing the fusion MLP in
Eq.~\eqref{eq:fusion-concat-projection}, and removing the SwiGLU FFN in
Eq.~\eqref{eq:swiglu-ffn}.

\begin{table}[t]
\centering
\caption{Study of different fusion strategies. Larger NE regression indicates a
larger quality drop relative to \model and therefore a more important design
choice.}
\label{tab:fusion-strategy-study}
\begin{tabular}{@{}llr@{}}
\toprule
Hierarchy & Fusion strategy & NE reg. (\%) \\
\midrule
Reference & \model & 0.00 \\
Architecture level & Shallow-hybrid fusion & 0.25 \\
Module level & Avg.-pooling fusion & 0.23 \\
Module level & Fusion w/o MLP projection & 0.11 \\
Module level & Fusion w/o SwiGLU FFN & 0.08 \\
\bottomrule
\end{tabular}
\end{table}

The results show that both levels of the fusion strategy matter. At the
architecture level, shallow-hybrid fusion performs worse than \model,
suggesting that progressive cross-layer fusion is important for jointly
leveraging non-sequence and sequence features. At the module level, replacing attention with average pooling leads to a clear
degradation, indicating that attention is important for allowing high-order
non-sequence feature interactions to query long-history representations and
retrieve relevant behavioral evidence during fusion. Removing the MLP projection
or the SwiGLU FFN also hurts performance, showing that learned cross-branch
integration and post-fusion nonlinear transformation both contribute to
recommendation quality.

\subsection{Online experiments}
\label{sec:online-experiments}

We further evaluate \model through online A/B tests in a large-scale social media recommendation
surface and compare it with the online baseline. The online A/B test runs for
14 days under the production serving stack, using the same candidate generation
and serving constraints, e.g., latency, as the online baseline.
On the platform, a 0.03\% win in the primary evaluation metric is considered
significant. Metric~1 and Metric~2 are online metrics supporting the primary-metric lift is directionally reliable.

To make \model practical for online deployment, the training QPS optimization
techniques described in Section~\ref{sec:efficiency-optimization} improve
training throughput by 30\%, while the inference QPS optimizations further
increase serving throughput by 22\%, against the unoptimized alternative. 

The online results are summarized in Table~\ref{tab:online-results}. Compared with the online baseline, \model improves the primary evaluation metric by 0.113\% while also increasing Metric~1 and Metric~2 by 0.824\% and 1.820\%, respectively. On the systems side, these gains come with a 5\% inference QPS regression against the online baseline, which remained acceptable under the online serving budget. These results show unifying Wukong-style feature-interaction modeling with HSTU-style long-range sequence modeling improves online recommendation outcomes.

\begin{table}[t]
\centering
\caption{Online A/B test results of \model. Positive values indicate improvements; negative inference QPS indicates the serving-throughput regression.}
\label{tab:online-results}
\begin{tabular*}{0.8\columnwidth}{@{\extracolsep{\fill}}lr@{}}
\toprule
Metric & Relative change \\
\midrule
Primary evaluation metric & +0.113\% \\
Metric 1 & +0.824\% \\
Metric 2 & +1.820\% \\
Inference QPS & -5\% \\
\bottomrule
\end{tabular*}
\end{table}
\section{Conclusion}

This work studies how two representative scaling paths in industrial
recommendation---Wukong for non-sequential feature interaction and HSTU for
sequential behavior modeling---can be brought together in a single deployable
architecture. Instead of using shallow coupling between the two branches, \model preserves both as active components and lets them exchange
information through layer-wise attention-based fusion.

Our experiments show this Wukong--HSTU unification improves over scaling
either backbone alone, benefits from longer behavior histories and larger model
capacity, and remains practical for online serving. These results suggest the next stage of scaling
industrial recommenders may come not only from making individual backbones
larger, but also from progressively unifying, at each layer, complementary
architectures for non-sequential feature interaction and sequential behavior
modeling.

\begin{acks}
The WHALE work reflects years of dedication from generations of ranking engineers across Meta's multiple organizations. It builds on the foundational work of Wukong and HSTU, and we are deeply grateful to their authors and contributors. Without their pioneering efforts, WHALE would not have been possible. This work also represents the collective effort of hundreds of people, and we would like to acknowledge the following contributors, listed alphabetically by last name: Nathan Berrebbi, Xianjie Chen, Huihui Cheng, Litao Deng, Qin Ding, Shilin Ding, Jun Du, Yiping Han, Daisy He, Zixia Hu, Rui Jian, Meilei Jiang, Dai Li, Han Li, Shen Li, Wei Li, Guandeng Liao, Hao Lin, Chihuang Liu, Chloe Liu, Meng Liu, Linjian Ma, Matt Ma, Romil Shah, Zhenyu Su, Jianhui Sun, Mengzhen Sun, Ruohan Sun, Wei Sun, Mert Terzihan, Jia Wang, Shengzhi Wang, Ye Wang, Haotian Wu, Junnan Wu, Yan Xia, Hong Yan, Yiling You, Qunshu Zhang, Zhang Zhang, Yiyang Zhao, Jin Zhou, Zhao Zhu.
\end{acks}

\bibliographystyle{ACM-Reference-Format}
\bibliography{references}

\end{document}